\documentclass{article}

\usepackage{arxiv}

\usepackage[utf8]{inputenc} 
\usepackage[T1]{fontenc}    
\usepackage{hyperref}       
\usepackage{url}            
\usepackage{booktabs}       
\usepackage{amsfonts}       
\usepackage{nicefrac}       
\usepackage{microtype}      
\usepackage{lipsum}		
\usepackage{graphicx}
\usepackage{natbib}
\usepackage{doi}
\usepackage{multirow}
\usepackage{threeparttable}
\usepackage{caption}
\usepackage{placeins}

\makeatletter
\AtBeginDocument{%
  \expandafter\renewcommand\expandafter\section\expandafter
    {\expandafter\@fb@secFB\section}%
  \newcommand\@fb@secFB{\FloatBarrier
    \gdef\@fb@afterHHook{\@fb@topbarrier \gdef\@fb@afterHHook{}}}%
  \g@addto@macro\@afterheading{\@fb@afterHHook}%
  \gdef\@fb@afterHHook{}%
}
\makeatother

\makeatletter
\AtBeginDocument{%
  \expandafter\renewcommand\expandafter\subsection\expandafter
    {\expandafter\@fb@secFB\subsection}%
}
\makeatother

\title{AI instructional agent improves student’s perceived learner control and learning outcome: empirical evidence from a randomized controlled trial}

\author{
    Fei Qin \\
	School of Education \\
	Tsinghua University\\
	Beijing, China \\
	\texttt{qinfei@tsinghua.edu.cn} \\
	\And
	Zhanxin Hao \\
    School of Education \\
	Tsinghua University\\
	Beijing, China \\
	\texttt{zhanxin\_hao@tsinghua.edu.cn} \\
	\AND
	Jifan Yu \\
	School of Education \\
	Tsinghua University \\
    Beijing, China \\
	\texttt{yujifan@tsinghua.edu.cn} \\
	\And
	Zhiyuan Liu \\
	Department of Computer Science and Technology \\
	Tsinghua University \\
    Beijing, China \\
	\texttt{liuzy@tsinghua.edu.cn} \\
	\And
	Yu Zhang\thanks{Corresponding author. Address: 417 Wennan Building, Tsinghua University, Beijing, China, 100084. Email: zhangyu2011@tsinghua.edu.cn}\\
	School of Education \\
	Tsinghua University \\
	Beijing, China \\
	\texttt{zhangyu2011@tsinghua.edu.cn} \\
}

\date{}

\begin{document}
\maketitle
\begin{abstract}
This study examines the impact of an AI instructional agent on students' perceived learner control and academic performance in a medium demanding course with lecturing as the main teaching strategy. Based on a randomized controlled trial, three instructional conditions were compared: a traditional human teacher, a self-paced MOOC with chatbot support, and an AI instructional agent capable of delivering lectures and responding to questions in real time. Students in the AI instructional agent group reported significantly higher levels of perceived learner control compared to the other groups. They also completed the learning task more efficiently and engaged in more frequent interactions with the instructional system. Regression analyzes showed that perceived learner control positively predicted post-test performance, with behavioral indicators such as reduced learning time and higher interaction frequency supporting this relationship. These findings suggest that AI instructional agents, when designed to support personalized pace and responsive interaction, can enhance both students' learning experience and learning outcomes.
\end{abstract}

\keywords{AI instructional agent; perceived learner control; academic performance; randomized controlled trial; personalized learning}

\section{Introduction}
Lecture-based instruction remains one of the most enduring and widely adopted modalities in formal education, particularly in higher education contexts (French \& Kennedy, 2017; Kay et al., 2019; Crawford \& Parsell, 2025). When well designed, lectures can foster a sense of academic community, provide sustained narrative, and offer students exposure to emerging research and enhance learning efficiency through structured content delivery (Blenkinsop et al., 2016; French \& Kennedy, 2017; Fulford \& Mahon, 2020). However, they are also frequently criticized for limited interactivity and misalignment with student-centered approaches, as students increasingly seek more flexible, self-directed, and active learning opportunities (Folley, 2010; Stearns, 2017). Although massive open online courses (MOOCs) have attempted to address these limitations through segmented, self-paced video instruction, their lack of dialogic responsiveness has constrained their pedagogical value, particularly in formal, credit-bearing education (Crawford \& Parsell, 2025; Yu et al., 2024). 

Artificial intelligence has been widely recognized for its potential to advance personalized learning, through adaptive feedback, customized learning paths, and dynamic content delivery (Ayeni et al., 2024). However, most current AI-enhanced learning environments have focused primarily on content automation, learner analytics, and management functionalities. Platforms such as Google Classroom, Seesaw, and Canvas are frequently described as AI-powered; however, their core functionality centers on resource organization and administrative support rather than instructional delivery (Perrotta et al., 2021; Yang-Heim \& Lin, 2024; Mpungose \& Khoza, 2022). Furthermore, AI conversational agents or chatbots integrated into these systems often lack pedagogical structure and operate reactively, providing limited support for sustained instructional engagement (McTear, 2022; Darvishi et al., 2024; Yusuf \& Daylamani-Zad, 2025). 

In contrast, the concept of an AI instructional agent, defined here as an AI system capable of delivering lecture content while simultaneously responding to students in real time, offers a more pedagogically grounded application. One of its distinguishing features is its potential to enable behavior-level learner control, allowing students to navigate, pause, and interact with instructional content in a more autonomous manner. This functionality is exemplified by the MAIC system (Massive AI-powered Course), which employs LLM-driven agents to deliver structured lectures while responding to real-time input from students through speech and multimodal content, thus simulating both instructional delivery and interactive tutoring (Yu et al., 2024; Hao et al., 2025).

Understanding the instructional implications of AI instructional agents is crucial, as these systems represent a pedagogically integrated approach that combines real-time content delivery with interactive responsiveness. When designed to deliver lectures dynamically while supporting learner-initiated interaction, AI instructional agents can help address the longstanding tension between scalability and individualization in classroom teaching (Wang et al., 2023; Wu et al., 2024). 

This study focuses on the construct of perceived learner control (PLC), which refers to the perceived ability of students to regulate the pace, sequence, and content of instruction (Kraiger \& Jerden, 2007; Karim \& Behrend, 2014). PLC has received extensive attention in the context of computer-based learning environments (Chou \& Liu, 2005; Sorgenfrei \& Smolnik, 2016), yet its manifestation in AI-mediated instructional settings remains underexplored. To address this gap, we conducted a randomized controlled trial comparing three instructional formats: (1) live instruction delivered by a human teacher, (2) a self-paced MOOC-style video lecture with chatbot support, and (3) an AI instructional agent embedded within our MAIC (Massive AI-empowered Courses) system. By keeping the content consistent across conditions and varying only the mode of instructional delivery, this study aims to isolate the effects of AI-mediated lecture delivery on students’ perceived learner control and learning outcomes.

\section{Literature review}
\label{sec:headings}

\subsection{Perceived Learner Control in Learning Environments}
Learner control refers to how much choice learners have in directing their own study process (Lin \& Hsieh, 2001; Kraiger \& Jerden, 2007). It has long been seen as a way to improve learning through greater autonomy, but decades of research show mixed outcomes. Some studies report that giving learners more control over elements like pace, sequence, or content can boost learning and motivation (Chou \& Liu, 2005; Corbalan, Kester \& Van Merriënboer, 2009), yet others find little benefit and even negative effects when novices learn without guidance (Hasler, Kersten \& Sweller, 2007; Al Nashrey, 2020). 

Meta-analyses reflect these mixed findings: overall, learner control yields only modest improvements in objective outcomes (Karich, Burns \& Maki, 2014; Hauk \& Gröschner, 2022). Nevertheless, it tends to enhance learners’ engagement and satisfaction (Karim \& Behrend, 2014; Brown, Howardson \& Fisher, 2016), indicating that students appreciate having choices even if performance gains are limited. 

The effectiveness of learner control depends on several factors. Learners with high prior knowledge or strong self-regulation skills benefit more from autonomy, whereas less-prepared learners often become overwhelmed by excessive choices (Hung et al., 2010; Hasler, Kersten \& Sweller, 2007). To address this, researchers have explored guided or structured approaches to support effective control. For instance, segmenting content so that students control pacing in manageable units provides autonomy with structure and can improve comprehension (Mayer et al., 2019). Hybrid models blending learner choice with system guidance (shared control) likewise aim to offer freedom without overloading the learner (Corbalan, Kester \& Van Merriënboer, 2009; Vandewaetere \& Clarebout, 2011). Additionally, “learner control” encompasses multiple dimensions (e.g., pace, sequence, content), which may have distinct effects in different settings (Karim \& Behrend, 2014).

In parallel, research has shifted toward perceived learner control — the learner’s subjective sense of autonomy. According to self-determined theory, the more autonomy students perceive, the more effort they put into their learning (Ryan \& Deci, 2020).Studies show that when students feel in control of their learning, their motivation and engagement increase (Buchem, Tur \& Hoelterhof, 2020; Lysne et al., 2023). Meaningful choices foster a sense of ownership, which is linked to greater persistence and satisfaction (Jung et al., 2019; Wang et al., 2025). Conversely, simply adding control features without adequate support may not raise learners’ perceived control if they feel unsure how to use the options. Recent work therefore emphasizes not just giving control, but also guiding learners to understand and capitalize on their autonomy (Ooge et al., 2025; Borchers et al., 2025). 

Research on learner control continues to evolve with new technologies. In online courses and MOOCs, giving learners adaptive paths or optional activities can influence their sense of control (Jung et al., 2019; Li, Wang \& Wallace, 2022). The rise of AI-driven learning environments raises new questions about balancing algorithmic guidance with learner agency (Brusilovsky, 2024). Initial studies suggest that well-implemented control features in AI tutoring systems can enhance the learning experience (Wambsganss et al., 2024). However, few studies have examined how AI-based instruction affects students’ perceived control compared to traditional teaching. This gap motivates the current study to examine whether an AI instructional agent can enhance students’ perceived control relative to a human instructor.

\subsection{AI-driven Instructional Agents}
AI instructional agents – often called pedagogical agents, conversational tutors, or virtual instructors – are intelligent software entities designed to emulate a teacher’s role in delivering content and guiding learning (Lane \& Schroeder, 2022; ji, Han \& Ko, 2023). These agents integrate features like natural-language dialogue, adaptive feedback, and multimedia presentation (Dever et al., 2023; Yusuf et al., 2025). For example, an AI agent might appear as an animated on-screen avatar delivering a lecture, or as a chatbot/voice tutor that answers learners’ questions. Critically, such agents actively structure and sequence instruction (mirroring human tutors), rather than merely serving as static content repositories (Yusuf et al., 2025; Ortega-Ochoa et al., 2024). 

In practice, instructional agents take diverse forms. One emphasizes structured content delivery, often in the form of AI-generated video lecturers or avatars. These agents simulate teacher presence and deliver prepared material but lack real-time interactivity (Arkün-Kocadere \& Özhan, 2024; Bai et al., 2025). Others function as interactive dialogue tutors embedded in learning platforms: students can pause or ask questions, and the agent responds in real time (Ding et al., 2023; Sánchez-Vera, 2025). For instance, Basri et al. (2023) created a mobile augmented-reality virtual tutor for medical training, and Sánchez-Vera (2025) examined a chatbot guiding students through exam prep. These two forms differ not only in interactivity but also in pedagogical emphasis. 

Studies generally report that well-designed agents yield moderate gains in learning and engagement. Meta-analyses find small-to-moderate effect sizes favoring adaptive pedagogical agents over passive materials (Davis et al., 2023; Siegle et al., 2023). In one randomized trial, medical trainees learning surgical skills from an AI tutor performed on par with those taught by expert surgeons (Fazlollahi et al., 2022). Similarly, Prasongpongchai et al. (2024) and Rathika et al. (2024) observed that AI-driven tutors effectively supported learning in financial and STEM topics. However, outcomes vary: when agents lack real-time interactivity or personalization, benefits often vanish. For example, Bai et al. (2025) found that changing an AI teacher’s appearance or voice did not significantly affect preschoolers’ learning from story videos, and Arkün-Kocadere and Özhan (2024) reported equal test scores between AI-lectured (with no real-time interactivity) and human-lectured classes (but lower engagement with the AI). 

Crucial differences in agent design help explain mixed results. Agents that display social and emotional behavior – for instance, an enthusiastic tone or encouraging feedback – tend to increase motivation and effort (Beege \& Schneider, 2023; Lang et al., 2024). In contrast, simply making an agent more realistic or human-like can backfire: Li et al. (2024) observed that photorealistic instructor videos drew more attention but reduced learning efficiency, perhaps due to cognitive load. Most studies find that interactivity matters most: Wu et al. (2024) showed that students achieved higher comprehension when the agent allowed frequent questions and on-demand answers, compared to a non-responsive presentation. In sum, the most effective agents combine embodiment (an avatar or presence) with adaptive scaffolding – such as pausing lessons to field questions or adjusting difficulty – aligning with established learning theories (Sikström et al., 2022; Dever et al., 2024). 

Several research gaps remain in the current literature on AI instructional agents. First, comparative trials are still limited: more randomized controlled trials (RCTs) are needed that directly compare AI agents with expert human teachers across diverse subjects and educational contexts (Arkün-Kocadere \& Özhan, 2024; Bai et al., 2025). Second, while content-delivery agents and real-time interactive tutors have been studied separately, there is a lack of empirical research on hybrid systems that integrate both structured instructional delivery and dynamic learner interaction—such as AI teachers capable of lecturing and adaptively responding to student input. This absence of holistic design limits our understanding of how to fully replicate or augment the instructional functions of human teachers. These gaps should be addressed through carefully designed empirical studies that reflect real-world instructional settings.

\subsection{The Present Study}
Based on the literature review, the current study was designed to investigate the effect of an AI instructional agent on university students’ perceived learner control and learning outcomes, using a randomized controlled trial. 

We focused on two primary research questions:

•	RQ1: \textit{Does the type of instructor (AI agent, human teacher, or MOOC-style learning) influence students’ perceived learner control and their learning outcomes?}
In particular, do students learning from an AI instructional agent report higher perceived control over their learning process, and do they achieve different post-test performance, compared to students in the human-taught or MOOC conditions?

•	RQ2: \textit{To what extent do students’ perceived learner control and related self-directed learning behaviors predict their academic performance?}
We examine whether students who feel more in control of their learning (regardless of condition) tend to perform better on a post-test. Additionally, we consider objective behavioral indicators of control (such as the number of questions students asked and the amount of time they spent on the learning task) to see if these mediate or explain the relationship between perceived control and performance.

By addressing RQ1, we test the hypothesis that the AI instructional agent, which allows more flexible pacing and questioning, will lead to greater perceived control (and possibly improved learning outcomes) relative to a conventional classroom or a basic online course. RQ2 allows us to probe the mechanism of any performance differences: if perceived control is indeed an important factor, we expect it to be positively associated with learning success. We also anticipate that in the AI instruction environment, students might exhibit specific behaviors (like frequent questioning or efficient time use) that reflect their exercise of control, and that those behaviors could correlate with better understanding of the material.

\section{Research Design}
\subsection{RCT Design}
\begin{figure}[h]
    \centering
    \includegraphics[width=0.5\linewidth]{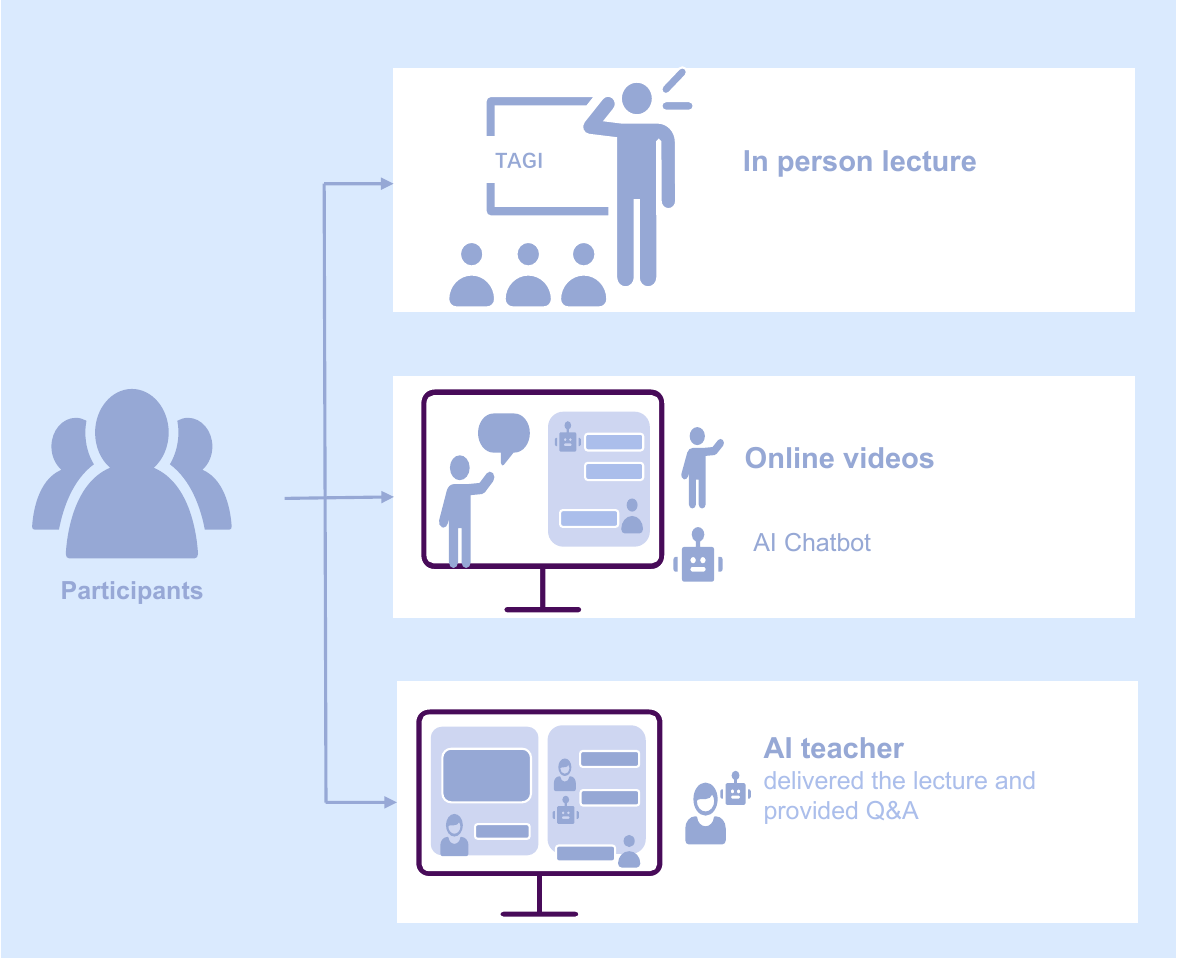}
    \caption{RCT Design Diagram}
    \label{fig:rct-design}
\end{figure}

This study employed a randomized controlled trial (RCT) design(see Figure \ref{fig:rct-design}). The core experimental group and control group were the AI instructional agent group and the human teacher group. In the human teacher condition, a senior instructor named “Teacher L.” from the Department of Computer and Science delivered in-person lectures in a traditional classroom setting. In the AI teacher group, instruction was provided via the MAIC platform (Yu et al., 2024), which integrates key instructional functions such as AI-generated lecture scripts, AI voice-based lecturing, and AI-agent-based interaction. During instruction, students were able to pause and interrupt the lecture to ask questions, which were answered in real time by the AI agent—also referred to in this study as “Teacher L.”

To ensure internal validity, several confounding variables were controlled across -groups. First, the instructional content was held constant. The same two modules from the general education course Towards General Artificial Intelligence were selected for all conditions. The instructional content was primarily focused on knowledge transmission and was of moderate difficulty. The PowerPoint slides and lecture scripts were identical across groups. While the MAIC platform has the capacity to autonomously generate lecture scripts, in this study the AI teacher used scripts used by the human instructor to ensure experimental consistency. Furthermore, the AI voice used for instruction was synthesized from the voice of the human teacher in the control condition, and the final version was approved by the instructor for naturalness and fidelity.

To better contrast the human teacher and AI teacher conditions, a third group—the MOOC group—was included as an intermediate instructional modality. In the MOOC group, students studied through pre-recorded lecture videos accompanied by an AI chatbot trained on the course content. However, unlike the AI teacher group, the chatbot was not named “Teacher L.”, nor was it integrated into the learning interface. Students had to separately invoke the chatbot for interaction. From a human-computer interaction perspective, this reduced ease of access and perceived continuity of the learning agent. In contrast, the AI teacher group experienced both instruction and interaction through the same AI identity, enhancing the perceived unity of the instructional agent.

In summary, the study involved three conditions: the human teacher group, the MOOC group, and the AI teacher group. All three conditions shared the same instructional content and audio characteristics.

\subsection{Sample}
Participants were recruited from universities in Beijing. A total of 140 undergraduate students participated in the study (63 males, 76 females), with an average age of 21.5 years old. This study was approved by the Institutional Review Board and all students provided informed consent before participating. Prior to the intervention, all participants completed a pre-test questionnaire that assessed demographic variables (e.g., gender, age), Big Five personality traits (O’Connor \& Paunonen, 2007; Zhang et al, 2019), self-regulated learning (Dowson \& McInerney, 2004; Wang \& Pomerantz, 2009), attitudes toward AI (Venkatesh, Thong, \& Xu, 2012), as well as a content-based knowledge pre-test.

Participants were randomly assigned to the three groups. No significant group difference was found in gender, age, personality traits, self-regulated learning, and pre-test scores.

\subsection{Measures}
The primary outcome variable in this study was perceived learner control. As no existing validated scale sufficiently captured learner control across both AI- and human-teacher learning contexts, a new scale was developed based on theoretical literature (see Table \ref{tab:pcl-measure}). After data collection and reliability and validity analysis, four items were retained. The Cronbach’s alpha was 0.823. An exploratory factor analysis using principal component analysis (KMO = 0.780; Bartlett’s test: $\chi^2$ = 212.684, df = 6, p < .001) identified a single factor from the four items. This factor accounted for $66.15\%$ of the total variance, with strong loadings (0.709–0.871), supporting the effectiveness of the scale.

\begin{table}[h]
    \caption{The Measurement of Perceived Learner Control}
    \centering
    \begin{tabular}{cc}
    \toprule
    Item 1     &  In this class, I am able to actively decide how to participate in classroom activities. \\
    Item 2     &  In this class, I feel that I can autonomously decide the pace of my learning. \\
    Item 3     &  In this class, I am able to control the way I learn. \\
    Item 4     &  In this class, I can decide how to allocate my time. \\
    \bottomrule
    \end{tabular}
    \label{tab:pcl-measure}
\end{table}

In addition to perceived learner control, we also measured student engagement using a classical three-dimensional scale encompassing cognitive, emotional, and behavioral engagement, which served as indicators of students’ process-oriented learning involvement (Reeve \& Tseng, 2011).

Beyond self-reported measures, behavioral data were collected to complement and contextualize perceived learner control. In the AI teacher condition, students learned at their own pace, making learning duration a confounding variable—students with lower learning abilities may use more time to learn, or longer learning duration may improve learning outcome, compared with human-teacher condition. Additionally, since the AI platform enabled real-time interactions, the frequency of student-initiated questions may reflect control-related behaviors, and should be controlled in regression. 

Lastly, academic performance was assessed through a post-test consisting of 22 multiple-choice questions (each with four options), designed by the course instructor. After item response analysis, 16 questions were included to measure the students’ posttest scores, which was computed as the proportion of correct answers.

\subsection{Data analysis}
Although 140 participants completed the experiment, due to concerns about data validity in self-report measures, responses in which over $90\%$ of scale items received identical ratings were excluded. This resulted in the removal of 15 responses. The final valid sample included 125 students: 41 in the human teacher group, 43 in the MOOC group, and 41 in the AI teacher group. 

For research question 1, to examine whether there were statistically significant differences in PLC, Posttest scores, learning engagement, across different groups, a one-way analysis of variance (ANOVA) was conducted. All the process and outcome variables were analyzed to compare the three conditions. Where significant group-level effects were identified, post-hoc pairwise comparisons were performed using Tukey’s Honestly Significant Difference (HSD) test to control for Type I error inflation.

For research question 2, firstly, the correlation of posttest scores with indicators such as PLC and learning duration, feedback numbers was analyzed to investigate whether perceived learner control predicts students’ academic performance, and the correlation between PLC and behavioral indicators, pre-intervention variable such as self-regulated learning. A multiple linear regression model was also employed. The dependent variable was the post-test score (posttest). The key independent variable of interest is the perceived learner control. Demographic variable and individual characteristics were included as controls: gender, age, self-regulated learning (SRL), and pre-test score (pretest).

Let $Y_i$  denote the post-test score of student $i$. The regression model is specified as follows:

\[
Y_i=\beta_0+\beta_1perceived~learner~control_i+\beta_2gender_i+\beta_3age_i+\beta_4srl_i+\beta_5pretest_i+\epsilon_i
\]

where:

$\beta_0$ is the intercept term; 
$\beta_1$ to $\beta_9$ are the regression coefficients for each predictor;
$\epsilon_i\sim N(0, \sigma^2)$ denotes the error term.

\section{Findings}
\subsection{AI Instructional agent Significantly Increases Students’ Perceived Learner Control and academic performance}
\begin{figure}[htpb]
    \centering
    \includegraphics[width=0.8\linewidth]{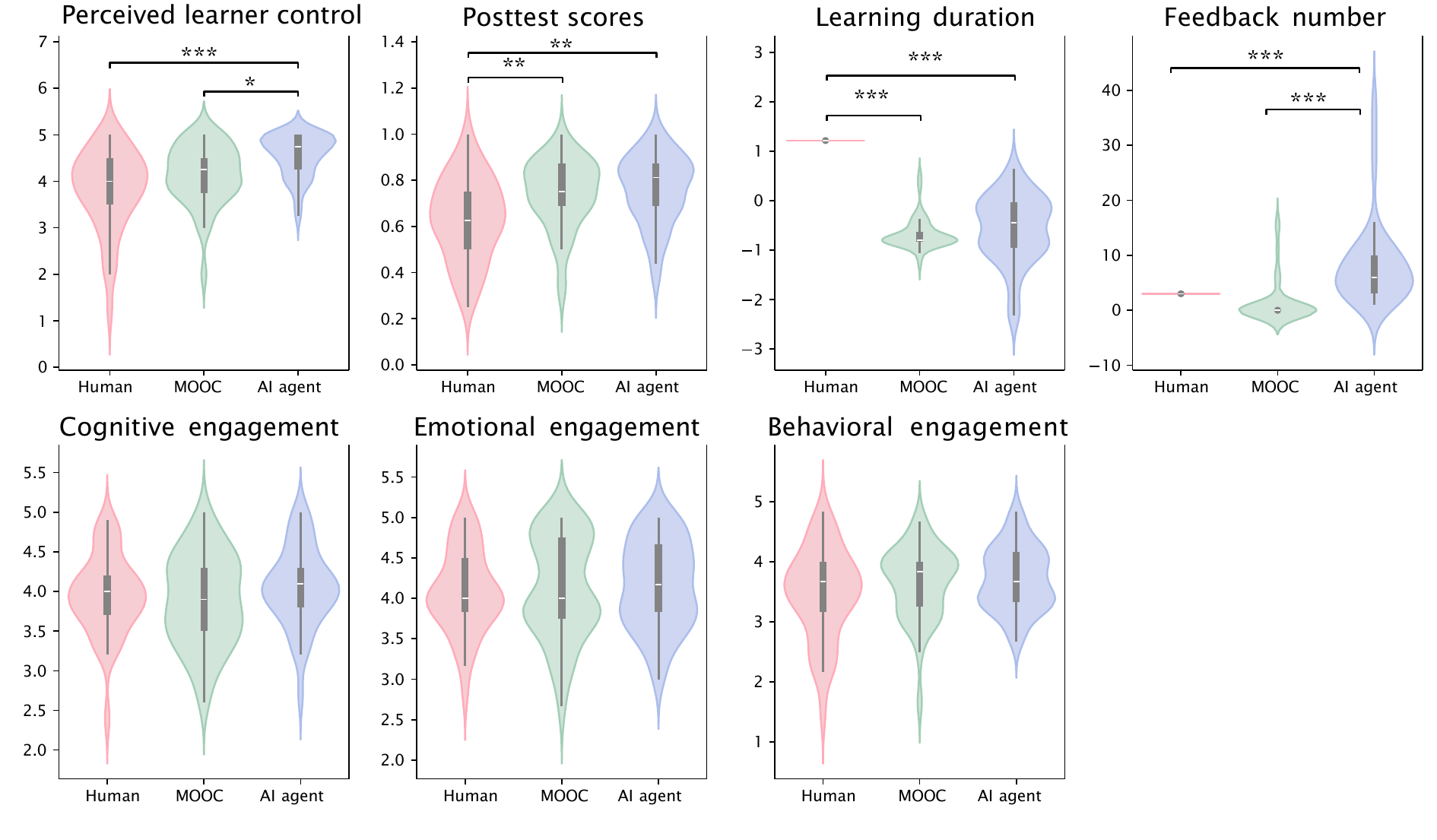}
    \caption{The Comparison between Three Groups of Key Variables}
    \label{fig:comp-three}

    \caption*{Note: * p < .05, ** p < .01, *** p < .001.}
\end{figure}

To examine whether students' perceived learner control differed across instructional contexts, a one-way ANOVA followed by Tukey’s HSD post-hoc tests was conducted on the three experimental groups: human teacher, AI agent, and MOOC-based learning. The results are summarized in Table \ref{tab:anova-hsd} and presented in Figure \ref{fig:comp-three}. 

Significant group differences were observed in the perceived learner control, F(2, 122) = 12.155, p < .001. Post-hoc comparisons revealed that students in the AI instructional group reported significantly higher levels of PLC compared to those in the human teacher group (M difference = 0.732, p < .001) and the MOOC group (M difference = 0.416, p < .05). This suggests that the AI-powered instructional setting, which allowed students to pause or accelerate lectures, ask questions in real time, and control the pacing of learning, significantly enhanced students' perceived capacity to self-direct their learning activities.

Additional analyses of behavioral variables revealed that the AI instructional agent group exhibited significantly higher interaction frequency than both the human teacher and MOOC groups, F(2, 122) = 21.755, p < .001, with a mean difference of 5.410 and 7.015 respectively. This indicates that students in the AI condition were more likely to initiate feedback-seeking behaviors during learning.
In contrast, both the AI and MOOC groups showed significantly shorter learning durations compared to the human teacher group, F(2, 122) = 265.168, p < .001, with reductions of approximately 30 minutes. While reduced time alone does not imply better learning, in the context of a self-paced environment with equivalent instructional content, it may reflect more efficient time usage and self-management by learners.

Importantly, these objective behavioral indicators—increased interaction frequency and reduced learning time—align with the observed group differences in perceived learner control, suggesting that students in the AI teacher condition not only reported feeling more in control, but also exhibited behavioral patterns consistent with that perception.

In addition, significant group differences were also found in post-test scores, F(2, 122) = 7.673, p < .01. Students in the AI group achieved significantly higher post-test scores than those in the human teacher group (M difference = 0.128, p < .01), and also outperformed the MOOC group (M difference = 0.112, p < .01). These results suggest that the instructional benefits of the AI agent extended beyond students’ subjective experience, contributing to improved learning outcomes as measured by objective academic performance.

Taken together, the findings indicate that AI-powered instruction meaningfully enhances students’ perceived learner control and promotes more interactive and efficient learning behaviors. The observed differences in both subjective ratings and behavioral indicators suggest that the AI-driven instruction environment more effectively supports learner-controlled learning.

\begin{table}[htbp]
    \centering
    \begin{threeparttable}
    \caption{Group Comparison: ANOVA and Tukey HSD Test}
    \begin{tabular}{lrlr}
    \toprule
    Variable                                   & F\_Value   & Group\_Comparison & Difference \\
    \midrule
Perceived learner control & 12.155$^{***}$  & AI-human          & 0.732$^{***}$   \\
                                           &            & AI-MOOC           & 0.416$^{*}$     \\
Cognitive engagement                       & 1.092      &                   &            \\
Emotional engagement                       & 0.215      &                   &            \\
Behavioral engagement                      & 1.670      &                   &            \\
Duration                                   & 265.168$^{***}$ & MOOC-human        & -1.960$^{***}$  \\
                                           &            & AI-human          & -1.860$^{***}$  \\
Feedback number                            & 21.755$^{***}$  & AI-human          & 5.410$^{***}$   \\
                                           &            & AI-MOOC           & 7.015$^{***}$   \\
Posttest scores                            & 7.673$^{**}$    & MOOC-human        & 0.112$^{**}$    \\
                                           &            & AI-human          & 0.128$^{**}$   \\
    \bottomrule
    \end{tabular}
    \label{tab:anova-hsd}
    \begin{tablenotes}
    \item Notes: * p < .05, ** p < .01, *** p < .001.
    \end{tablenotes}
    \end{threeparttable}
\end{table}

\subsection{Perceived Learner Control Positively Predicts Academic Performance}
\begin{figure}[htbp]
    \centering
    \includegraphics[width=0.8\linewidth]{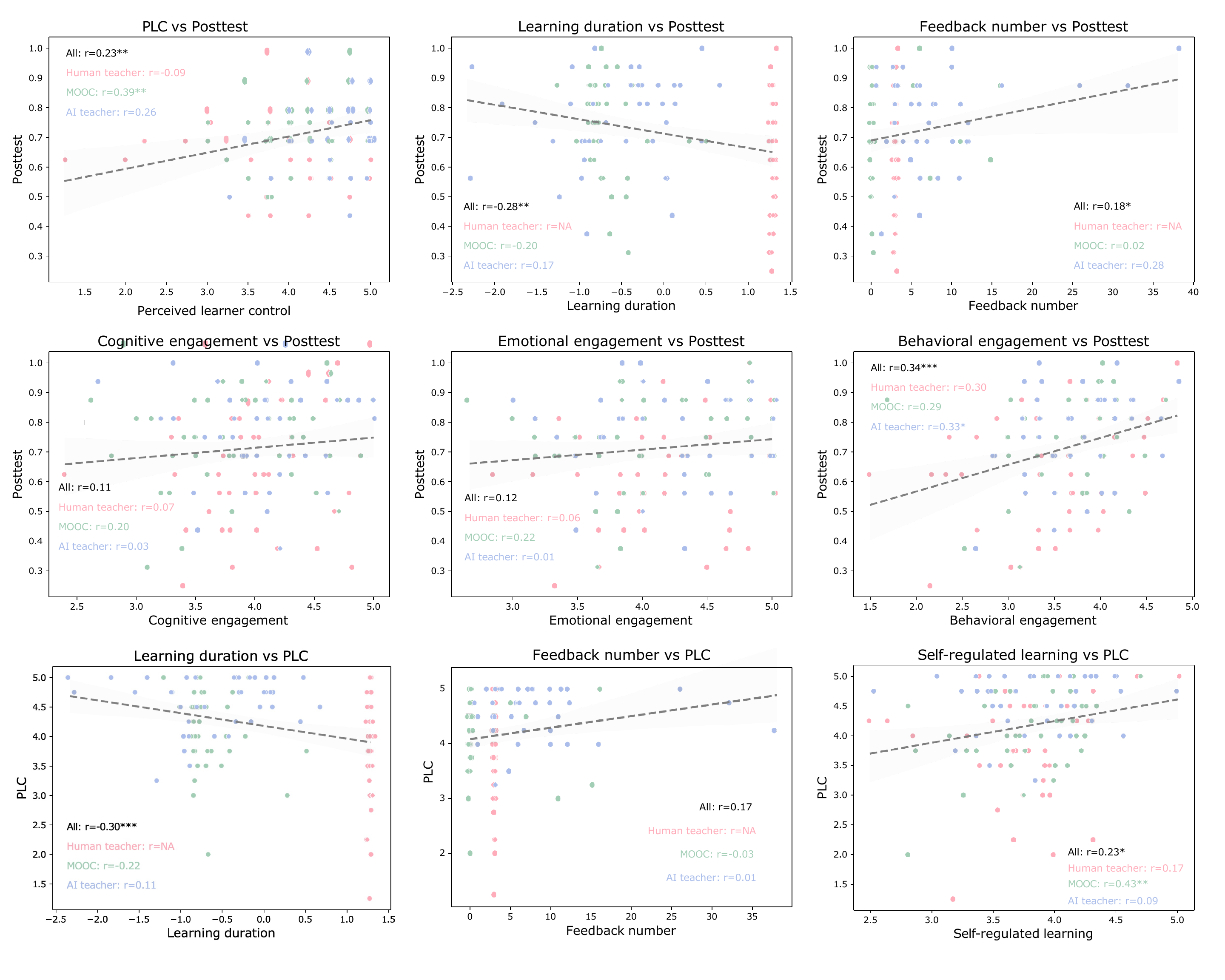}
    \caption{The Correlation between Posttest and Other Variables}
    \label{fig:corr-post}
\end{figure}

To assess whether PLC predicts students’ academic performance, bivariate correlation analyses were conducted between post-test scores and key variables (see Figure \ref{fig:corr-post}). Post-test scores were significantly correlated with PLC (r = 0.23, p < .01), learning duration (r = –0.28, p < .01), feedback frequency (r = 0.18, p < .05), and behavioral engagement (r = 0.34, p < .001), suggesting links between subjective perception, behavioral regulation, and learning outcomes. However, the posttest scores show no significant correlation with cognitive and emotional engagement.

Further analyses revealed that PLC was negatively correlated with learning duration (r = –0.30, p < .001), and positively correlated with self-regulated learning (r = 0.23, p < .05), while no significant correlation was found with feedback frequency. Since SRL was measured at the pretest stage, this result suggests that students with stronger pre-existing self-regulated learning abilities were more likely to perceive a higher degree of control during the learning process. The negative association with learning duration indicates that students who reported greater perceived control tended to complete the task more efficiently, reflecting more purposeful time management.

To further investigate the predictive role of PLC, a multiple linear regression model was estimated with post-test scores as the dependent variable. PLC was included as the key predictor, along with gender, age, self-regulated learning (SRL), and pre-test score as controls. The results showed that PLC significantly predicted post-test performance ($\beta$ = 0.055, p < .05), indicating that students who perceived greater control over their learning tended to perform better, even after accounting for prior achievement and learner background. Pre-test scores were also significant ($\beta$ = 0.223, p < .05); all other covariates were non-significant. This model was estimated with a statistical power of 0.948, suggesting sufficient sensitivity to detect small-to-medium effect sizes.

Taken together, these results confirm that perceived learner control positively contributes to academic achievement and is reflected in students’ behavioral indicators. The evidence suggests that control is both felt and enacted—perceived through subjective experience and expressed through behavioral decisions like efficient time use and purposeful interaction.

\section{Conclusion and Discussions}
This study examined the instructional effects of an AI-driven agent capable of delivering real-time lecture content, with a focus on students’ perceived learner control (PLC) and learning outcomes. The findings demonstrate that when AI agents are designed not merely as conversational tools but as integrated instructional systems, they can meaningfully enhance students’ sense of control and support academic performance.

Students in the AI instructional agent group reported significantly higher PLC and achieved better learning outcomes than those in the human teacher group, with similar advantages over the MOOC group in terms of perceived control. These results highlight that instructional effectiveness depends on how control and interaction are embedded into the delivery process, given the same content quality. Prior work has shown that while learner control can foster motivation and satisfaction (Karim \& Behrend, 2014; Buchem, Tur \& Hoelterhof, 2020), its benefits are conditional on structure and guidance (Hasler, Kersten \& Sweller, 2007; Hung et al., 2010). The AI agent in this study provided structured content delivery along with flexible pacing, pausing, and questioning features—an integration that enabled learners to exercise autonomy without the disorientation often associated with unstructured freedom.

The behavioral data supported this interpretation. Students in the AI group interacted more frequently with the system and completed the task more efficiently, suggesting that the agent’s design facilitated focused, self-paced learning. These outcomes align with findings that learner-centered pacing, when scaffolded within a coherent instructional flow, improves both engagement and comprehension (Mayer et al., 2019; Corbalan, Kester \& Van Merriënboer, 2009). Furthermore, regression analyses confirmed that PLC significantly predicted post-test performance, consistent with theories emphasizing the motivational and cognitive value of perceived autonomy (Ryan \& Deci, 2020; Jung et al., 2019; Wang et al., 2025).

Notably, our findings contrast with studies on non-interactive lecture agents that show limited gains despite high embodiment (Arkün-Kocadere \& Özhan, 2024; Bai et al., 2025). By enabling real-time interaction, our AI agent not only increased behavioral engagement—evidenced by more frequent interactions and shorter completion time—but also facilitated efficient, self-paced learning. These patterns align with the theoretical proposition that perceived autonomy is both felt and enacted (Jung et al., 2019; Buchem et al., 2020).

In light of recent calls for more holistic and context-aware agent design (Brusilovsky, 2024), our findings offer timely evidence that AI agents can move beyond tool-like support roles to take on core instructional functions. By simulating aspects of teacher behavior—sequencing content, responding to learners, and guiding attention—the agent demonstrates how AI can complement or even enhance traditional teaching in structured settings.

Despite the encouraging results, it is important to acknowledge the limitations of this study. First, the experiment was conducted in a university general education course with moderate content difficulty and a lecture-based instructional approach. The course did not involve high-level challenge, problem-based learning (PBL), or sustained dialogic teaching. Therefore, the findings may not generalize to other course types or pedagogical models that emphasize deeper reasoning, inquiry-based learning, or extensive interaction. Second, the study focused on immediate learning outcomes (post-test administered right after the intervention). It remains unclear whether the AI instructional agent supports long-term retention or knowledge transfer—an important direction for future research. It would also be valuable to examine whether the heightened sense of control observed in the short term leads to improved study habits or conceptual understanding over time.

The present study is expected to offer a helpful basis for future investigations. As AI instructional agents continue to evolve, it will be important to anchor their development in sound pedagogical theory and evidence-based instructional design, ensuring that technological innovation serves meaningful educational goals. Further randomized controlled trials in varied disciplinary settings, with diverse student populations and instructional approaches, will be instrumental in deepening our understanding of how, when, and for whom such systems are most effective. Through such efforts, AI-powered instructional agents may move closer to becoming truly transformative tools in education.

\section*{References}

Al Nashrey, A. A. (2020). Does Giving the Learner Control Improve Learner Success? International Journal of Education, 12(3), 135. 

Arkün-Kocadere, S. \& Özhan, Ş. Ç. (2024). Video Lectures With AI-Generated Instructors: Low Video Engagement, Same Performance as Human Instructors. International Review of Research in Open and Distributed Learning, 25(3), 350–369. 

Ayeni, O. O., Al Hamad, N. M., Chisom, O. N., Osawaru, B., \& Adewusi, O. E. (2024). AI in education: A review of personalized learning and educational technology. GSC Advanced Research and Reviews, 18(2), 261-271.

Bai, J., Cheng, X., Zhang, H., Qin, Y., Xu, T., \& Zhou, Y. (2025). Can AI-generated pedagogical agents (AIPA) replace human teacher in picture book videos? The effects of appearance and voice of AIPA on children’s learning. Education and Information Technologies. Scopus. 

Basri, N. A. H., Rahmat, R. W. O. K., Madzin, H., \& Hod, R. (2023). A User Study on Virtual Tutor Mobile Augmented Reality from Undergraduate Medical Student Perspectives. Asia-Pacific Journal of Information Technology and Multimedia, 12(1), 139–154. Scopus. 

Beege, M., Nebel, S., Rey, G. D., \& Schneider, S. (2024). Learning with pedagogical agents in digital environments. In A. Gegenfurtner \& I. Kollar (Eds.), Designing Effective Digital Learning Environments (pp. 109–122). Routledge. 

Blenkinsop, S., Nolan, C., Hunt, J., Stonehouse, P., \& Telford, J. (2016). The lecture as experiential education: The cucumber in 17th-century Flemish art. Journal of Experiential Education, 39(2), 101-114.

Borchers, C., Ooge, J., Peng, C., \& Aleven, V. (2025). How Learner Control and Explainable Learning Analytics About Skill Mastery Shape Student Desires to Finish and Avoid Loss in Tutored Practice. Proceedings of the 15th International Learning Analytics and Knowledge Conference, 810–816. 

Brown, K. G., Howardson, G., \& Fisher, S. L. (2016). Learner Control and e-Learning: Taking Stock and Moving Forward. Annual Review of Organizational Psychology and Organizational Behavior, 3(1), 267–291. 

Brusilovsky, P. (2024). AI in Education, Learner Control, and Human-AI Collaboration. International Journal of Artificial Intelligence in Education, 34(1), 122–135. 

Buchem, I., Tur, G., \& Hoelterhof, T. (2020). The role of learner control and psychological ownership for self-regulated learning in technology-enhanced learning designs. Differences in e-portfolio use in higher education study programs in Germany and Spain.

Chou, S., \& Liu, C. (2005). Learning effectiveness in a Web‐based virtual learning environment: A learner control perspective. Journal of Computer Assisted Learning, 21(1), 65–76. 

Corbalan, G., Kester, L., \& J.G. Van Merriënboer, J. (2009). Dynamic task selection: Effects of feedback and learner control on efficiency and motivation. Learning and Instruction, 19(6), 455–465. 

Crawford, J., \& Parsell, M. (2025). Lectures in higher education: A 22-year systematic review. Journal of Applied Learning and Teaching, 8(1).

Davis, R. O., Park, T., \& Vincent, J. (2023). A Meta-Analytic Review on Embodied Pedagogical Agent Design and Testing Formats. Journal of Educational Computing Research, 61(1), 30–67. Scopus. 

Dever, D. A., Sonnenfeld, N. A., Wiedbusch, M. D., Schmorrow, S. G., Amon, M. J., \& Azevedo, R. (2023). A complex systems approach to analyzing pedagogical agents’ scaffolding of self-regulated learning within an intelligent tutoring system. Metacognition and Learning, 18(3), 659–691. 

Dever, D. A., Wiedbusch, M. D., Romero, S. M., \& Azevedo, R. (2024). Investigating pedagogical agents’ scaffolding of self-regulated learning in relation to learners’ subgoals. British Journal of Educational Technology, 55(4), 1290–1308. Scopus. 

Ding, L., Li, T., Jiang, S., \& Gapud, A. (2023). Students’ perceptions of using ChatGPT in a physics class as a virtual tutor: Revista de Universidad y Sociedad del Conocimiento. International Journal of Educational Technology in Higher Education, 20(1), 63. Publicly Available Content Database; Research Library; SciTech Premium Collection; Social Science Premium Collection. 

Dowson, M., \& McInerney, D. M. (2004). The development and validation of the Goal Orientation and Learning Strategies Survey (GOALS-S). Educational and Psychological Measurement, 64(2), 290-310.

Fazlollahi, A. M., Bakhaidar, M., Alsayegh, A., Yilmaz, R., Winkler-Schwartz, A., Mirchi, N., Langleben, I., Ledwos, N., Sabbagh, A. J., Bajunaid, K., Harley, J. M., \& Del Maestro, R. F. (2022). Effect of Artificial Intelligence Tutoring vs Expert Instruction on Learning Simulated Surgical Skills Among Medical Students: A Randomized Clinical Trial. JAMA Network Open, 5(2), e2149008. 

Fink, M. C., Robinson, S. A., \& Ertl, B. (2024). AI-based avatars are changing the way we learn and teach: Benefits and challenges. Frontiers in Education, 9. 

Folley, D. (2010). The lecture is dead, long live the e-lecture. Electronic Journal of e-Learning, 8(2), 93-100.

French, S., \& Kennedy, G. (2017). Reassessing the value of university lectures. Teaching in Higher Education, 22(6), 639654. 

Fulford, A., \& Mahon, Á. (2020). A philosophical defence of the university lecture. Oxford Review of Education, 46(3), 363-374.

Hao, Z., Jiang, J., Yu, J., Liu, Z., \& Zhang, Y. (2025). Student engagement in collaborative learning with AI agents in an LLM-empowered learning environment: A cluster analysis (No. arXiv:2503.01694). arXiv. 

Hasler, B. S., Kersten, B., \& Sweller, J. (2007). Learner control, cognitive load and instructional animation. Applied Cognitive Psychology, 21(6), 713–729. 

Hauk, D., \& Gröschner, A. (2022). How effective is learner-controlled instruction under classroom conditions? A systematic review. Learning and Motivation, 80, 101850. 

Herbert, C., \& Dołżycka, J. D. (2024). Teaching online with an artificial pedagogical agent as a teacher and visual avatars for self-other representation of the learners. Effects on the learning performance and the perception and satisfaction of the learners with online learning: Previous and new findings. Frontiers in Education, 9. 

Hung, M.-L., Chou, C., Chen, C.-H., \& Own, Z.-Y. (2010). Learner readiness for online learning: Scale development and student perceptions. Computers \& Education, 55(3), 1080–1090. 

Ji, H., Han, I., \& Ko, Y. (2023). A systematic review of conversational AI in language education: Focusing on the collaboration with human teachers. Journal of Research on Technology in Education, 55(1), 48–63. 

Jung, E., Kim, D., Yoon, M., Park, S., \& Oakley, B. (2019). The influence of instructional design on learner control, sense of achievement, and perceived effectiveness in a supersize MOOC course. Computers \& Education, 128, 377–388. 

Karich, A. C., Burns, M. K., \& Maki, K. E. (2014). Updated Meta-Analysis of Learner Control Within Educational Technology. Review of Educational Research, 84(3), 392–410. 

Karim, M. N., \& Behrend, T. S. (2014). Reexamining the Nature of Learner Control: Dimensionality and Effects on Learning and Training Reactions. Journal of Business and Psychology, 29(1), 87–99. 

Kay, R., MacDonald, T., \& DiGiuseppe, M. (2019). A comparison of lecture-based, active, and flipped classroom teaching approaches in higher education. Journal of Computing in Higher Education, 31, 449-471.

Kraiger, K., \& Jerden, E. (2007). A meta-analytic investigation of learner control: Old findings and new directions. S. M. Fiore \& E. Salas, Toward a science of distributed learning. (65–90). American Psychological Association. 

Lane, H. C., \& Schroeder, N. L. (2022). Pedagogical agents. In The handbook on socially interactive agents: 20 years of research on embodied conversational agents, intelligent virtual agents, and social robotics volume 2: Interactivity, platforms, application (pp. 307-330).

Li, L., Wang, X., \& Wallace, M. P. (2022). I determine my learning path, or not? A study of different learner control conditions in online video-based learning. Frontiers in Psychology, 13, 973758. 

Li, W., Wang, F., \& Mayer, R. E. (2024). Increasing the realism of on-screen embodied instructors creates more looking but less learning. British Journal of Educational Psychology, 94(3), 759–776. 

Lin, B., \& Hsieh, C. (2001). Web-based teaching and learner control: A research review. Computers \& Education, 37(3–4), 377–386. 

Lysne, D. A., De Caro-Barek, V., Støckert, R., Røren, K. A. F., Solbjørg, O. K., \& Nykvist, S. S. (2023). Students’ motivation and ownership in a cross-campus and online setting. Frontiers in Education, 8, 1062767. 

Mayer, R. E., Wells, A., Parong, J., \& Howarth, J. T. (2019). Learner control of the pacing of an online slideshow lesson: Does segmenting help? Applied Cognitive Psychology, 33(5), 930–935. 

McTear, M. (2022). Conversational ai: Dialogue systems, conversational agents, and chatbots. Springer Nature.

Mpungose, C. B., \& Khoza, S. B. (2022). Postgraduate students’ experiences on the use of Moodle and Canvas learning management system. Technology, Knowledge and Learning, 27(1), 1-16.

O’Connor, M. C., \& Paunonen, S. V. (2007). Big Five personality predictors of post-secondary academic performance. Personality and Individual differences, 43(5), 971-990.

Ooge, J., Vanneste, A., Szymanski, M., \& Verbert, K. (2025). Designing Visual Explanations and Learner Controls to Engage Adolescents in AI-Supported Exercise Selection. Proceedings of the 15th International Learning Analytics and Knowledge Conference, 1–12. 

Ortega-Ochoa, E., Arguedas, M., \& Daradoumis, T. (2024). Empathic pedagogical conversational agents: A systematic literature review. British Journal of Educational Technology, 55(3), 886–909.

Perrotta, C., Gulson, K. N., Williamson, B., \& Witzenberger, K. (2021). Automation, APIs and the distributed labour of platform pedagogies in Google Classroom. Critical Studies in Education, 62(1), 97-113.

Prasongpongchai, T., Pataranutaporn, P., Kanapornchai, C., Lapapirojn, A., Ouppaphan, P., Winson, K., … \& Maes, P. (2024). Interactive AI-Generated Virtual Instructors Enhance Learning Motivation and Engagement in Financial Education. In Proceedings of the International Conference on Artificial Intelligence in Education (Vol. 2151, pp. 217–225). Springer. 

Rathika, P., Yamunadevi, S., Ponni, P., Parthipan, V., \& Anju, P. (2024). Developing an AI-Powered Interactive Virtual Tutor for Enhanced Learning Experiences. International Journal of Computational and Experimental Science and Engineering, 10(4), 1594–1600. Scopus. 

Reeve, J., \& Tseng, C.-M. (2011). Agency as a fourth aspect of students’ engagement during learning activities. Contemporary Educational Psychology, 36(4), 257–267. 

Ryan, R. M., \& Deci, E. L. (2020). Intrinsic and extrinsic motivation from a self-determination theory perspective: Definitions, theory, practices, and future directions. Contemporary Educational Psychology, 61, 101860. 

Sánchez-Vera, F. (2025). Subject-Specialized Chatbot in Higher Education as a Tutor for Autonomous Exam Preparation: Analysis of the Impact on Academic Performance and Students’ Perception of Its Usefulness. Education Sciences, 15(1), Article 26. 

Siegle, R. F., Schroeder, N. L., Lane, H. C., \& Craig, S. D. (2023). Twenty-five Years of Learning with Pedagogical Agents: History, Barriers, and Opportunities. TechTrends, 67(5), 851–864. 

Sikström, P., Valentini, C., Sivunen, A., \& Kärkkäinen, T. (2022). How pedagogical agents communicate with students: A two-phase systematic review. Computers \& Education, 188, 104564. 

Sorgenfrei, C., \& Smolnik, S. (2016). The Effectiveness of E‐Learning Systems: A Review of the Empirical Literature on Learner Control. Decision Sciences Journal of Innovative Education, 14(2), 154–184. 

Stearns, S. (2017). What is the place of lecture in student learning today?. Communication Education, 66(2), 243-245.

Vandewaetere, M., \& Clarebout, G. (2011). Can instruction as such affect learning? The case of learner control. Computers \& Education, 57(4), 2322–2332. 

Venkatesh, V., Thong, J. Y., \& Xu, X. (2012). Consumer acceptance and use of information technology: extending the unified theory of acceptance and use of technology. MIS quarterly, 157-178.

Wambsganss, T., Benke, I., Maedche, A., Koedinger, K., \& Käser, T. (2024). Evaluating the Impact of Learner Control and Interactivity in Conversational Tutoring Systems for Persuasive Writing. International Journal of Artificial Intelligence in Education. 

Wang, L., De Vetten, A., Admiraal, W., \& Van Der Rijst, R. (2025). Relationship between perceived learner control and student engagement in various study activities in a blended course in higher education. Education and Information Technologies, 30(2), 2463–2484. 

Wang, Q., \& Pomerantz, E. M. (2009). The motivational landscape of early adolescence in the United States and China: A longitudinal investigation. Child Development, 80(4), 1272-1287.

Wang, Y., Gong, S., Cao, Y., \& Fan, W. (2023). The power of affective pedagogical agent and self-explanation in computer-based learning. Computers and Education, 195, 104723. 

Wu, C., Li, Y., Yang, H., Wang, X., Li, X., \& Jing, B. (2024). The Greater the Interaction, the Stronger the Learning Performance? Examining Pedagogical Agents’ Interactive Presence in Instructional Videos. Applied Cognitive Psychology, 38(5), 1304–1318.

Yang-Heim, G. Y. A., \& Lin, X. (2024). Teacher Candidates' Perspectives on the Integration of Digital Tools in Teacher Training Programs: A Case Study of Using Seesaw. International Journal of Technology-Enhanced Education (IJTEE), 3(1), 1-19.

Yu, J., Zhang, Z., Zhang-li, D., Tu, S., Hao, Z., Li, R. M., Li, H., Wang, Y., Li, H., Gong, L., Cao, J., Lin, J., Zhou, J., Qin, F., Wang, H., Jiang, J., Deng, L., Zhan, Y., Xiao, C., … Sun, M. (2024). From MOOC to MAIC: Reshaping Online Teaching and Learning through LLM-driven Agents (No. arXiv:2409.03512). arXiv. 

Yusuf, H., Money, A., \& Daylamani-Zad, D. (2025). Pedagogical AI conversational agents in higher education: a conceptual framework and survey of the state of the art. Educational technology research and development, 1-60.

Zhang, X., Wang, M.-C., He, L., Jie, L., \& Deng, J. (2019). The development and psychometric evaluation of the Chinese Big Five Personality Inventory-15. PLOS ONE, 14(8), e0221621. https://doi.org/10.1371/journal.pone.0221621

\end{document}